\documentclass[12pt]{article}
\newcommand{\bq}{\begin{equation}}
\newcommand{\eq}{\end{equation}}
\newcommand{\ba}{\begin{eqnarray}}
\newcommand{\ea}{\end{eqnarray}}

\newcommand{\nl }{ \nonumber  }

\newcommand{\p}{\partial}
\newcommand{\h}{\hspace{.5cm}}
\newcommand{\s}{\sigma}

\begin{document}
\vspace*{.5cm}
\begin{center}
{\bf STRING SOLUTIONS IN GENERAL BACKGROUNDS \vspace*{0.5cm}\\ P. Bozhilov}
\\ {\it Institute for Nuclear Research and Nuclear Energy, \\
Bulgarian Academy of Sciences, \\ 1784 Sofia, Bulgaria\\
E-mail:} bozhilov@inrne.bas.bg
\end{center}
\vspace*{0.5cm}

Motivated by the recent interest in the different aspects of the string/field 
theory duality, we describe an approach for obtaining exact string solutions 
in general backgrounds, based on two types of string embedding, allowing for
separation of the worldsheet variables $\tau$ and $\s$.

\vspace*{.5cm} {\bf PACS:} 11.25.-w, 11.27.+d, 11.30.-j

\vspace*{.5cm} {\bf Keywords:} Bosonic strings, AdS-CFT correspondence, 
space-time symmetries, integrable equations in physics. 

\vspace*{.5cm}

\section{Introduction}
Recently, there is a lot of interest in the investigation of the existing 
connections between different classical string configurations, their 
semi-classical quantization and the relevant objects in the dual gauge theories,
as well as between the corresponding integrable models appearing on the string
and field theory sides (see e.g. \cite{4} - \cite{89} \footnote{For previously
obtained string solutions in curved space-times, see e.g. 
\cite{NSetall,PB01} and the references in \cite{PB01} and \cite{NS03}.} 
and the references therein). In this connection, it 
seems useful to formulate an approach, which will allow us to obtain exact 
string solutions in general enough string theory backgrounds. Here, we 
describe such an approach, based on two types of string embedding, which allow 
for separation of the worldsheet variables $\tau$ and $\s$.

\setcounter{equation}{0}
\section{Exact string solutions in general backgrounds}
In our further considerations, we will use the Polyakov type
action for the bosonic string in a $D$-dimensional curved
space-time with metric tensor $g_{MN}(x)$, interacting with a
background 2-form gauge field $b_{MN}(x)$ via Wess-Zumino term
\ba\label{pa} &&S^{P}=\int d^{2}\xi\mathcal{L}^P,\h \mathcal{L}^P
=-\frac{1}{2}\left(T\sqrt{-\gamma}\gamma^{mn}G_{mn}-Q\varepsilon^{mn}
B_{mn}\right),\\ \nl && \xi^m=(\xi^0,\xi^1)=(\tau,\s),\h m,n =
(0,1),\ea where  \ba\nl &&G_{mn}= \p_m X^M\p_n X^N g_{MN},\h
B_{mn}=\p_{m}X^{M}\p_{n}X^{N} b_{MN}, \\ \nl &&(\p_m=\p/\p\xi^m,\h
M,N = 0,1,\ldots,D-1),\ea are the fields induced on the string
worldsheet, $\gamma$ is the determinant of the auxiliary
worldsheet metric $\gamma_{mn}$, and $\gamma^{mn}$ is its inverse.
The position of the string in the background space-time is given
by $x^M=X^M(\xi^m)$, and $T=1/2\pi\alpha'$, $Q$ are the string
tension and charge, respectively. If we consider the action
(\ref{pa}) as a bosonic part of a supersymmetric one, we have to
put $Q=\pm T$. In what follows, $Q =T$.

The equations of motion for $X^M$ following from (\ref{pa}) are:
\ba \label{em}
&&-g_{LK}\left[\p_m\left(\sqrt{-\gamma}\gamma^{mn}\p_nX^K\right) +
\sqrt{-\gamma}\gamma^{mn}\Gamma^K_{MN}\p_m X^M\p_n X^N\right]\\
\nl &&=\frac{1}{2}H_{LMN}\epsilon^{mn}\p_m X^M\p_n X^N,\ea where
\ba\nl
&&\Gamma_{L,MN}=g_{LK}\Gamma^K_{MN}=\frac{1}{2}\left(\p_Mg_{NL}
+\p_Ng_{ML}-\p_Lg_{MN}\right),\\ \nl &&H_{LMN}= \p_L b_{MN}+ \p_M
b_{NL} + \p_N b_{LM},\ea are the components of the symmetric
connection corresponding to the metric $g_{MN}$, and the field
strength of the gauge field $b_{MN}$ respectively. The constraints
are obtained by varying the action (\ref{pa}) with respect to
$\gamma_{mn}$: \ba\label{oc} \delta_{\gamma_{mn}}S^P=0\Rightarrow
\left(\gamma^{kl}\gamma^{mn}-2\gamma^{km}\gamma^{ln}\right)G_{mn}=0.\ea

In solving the equations of motion (\ref{em}) and constraints
(\ref{oc}), we will use the worldsheet gauge
$\gamma_{mn}=constants$. This will allow us to consider the
tensionless limit $T \to 0$, corresponding to small t'Hooft
coupling $\lambda \to 0$ on the field theory side. Of course, we
can always set $\gamma^{mn}=\eta^{mn}=diag(-1,1)$, to turn to the
usually used {\it conformal gauge}.

We will investigate the string dynamics in the framework of the
following two types of embedding: \ba\label{tGA} X^\mu (\tau,
\sigma)= \Lambda^\mu_0\tau + \Lambda^\mu_1\sigma + Y^\mu (\tau),\h
X^a (\tau, \sigma)= Y^a (\tau);\\ \label{sGA} X^\mu (\tau,
\sigma)= \Lambda^\mu_0\tau + \Lambda^\mu_1\sigma + Z^\mu
(\sigma),\h X^a (\tau, \sigma)= Z^a (\sigma);\\ \nl \Lambda^\mu_m
= const,\h (m=0,1).\ea Here, the embedding coordinates
$X^M(\tau,\s)$ are divided into $X^M=(X^\mu,X^a)$, where
$X^\mu(\tau,\s)$ correspond to the space-time coordinates $x^\mu$,
on which the background fields do not depend \ba\label{ob} \p_\mu
g_{MN} =0,\h \p_\mu b_{MN} =0.\ea In other words, we suppose that
there exist $n_\mu$ commuting Killing vectors $\p/\p x^\mu$, where
$n_\mu$ is the number of the coordinates $x^\mu$. In this case,
the ansatzes (\ref{tGA}) and (\ref{sGA}) allow for separation of
the variables $\tau$ and $\s$. By using (\ref{tGA}), one obtains
$\tau$-dependent dynamics, while by using (\ref{sGA}), one obtains
$\s$-dependent dynamics.

\subsection{$\tau$-dependent dynamics}
Here, we are going to use the ansatz (\ref{tGA}) for the string
embedding coordinates. In addition, we assume that the conditions
(\ref{ob}) on the background fields hold.

As far as only two of the constraints (\ref{oc}) are independent, we have
to choose consistently two of them. Our {\it independent} constraints,
with which we will work in this subsection, are given by
\ba\label{0}
\gamma^{00}G_{00}-\gamma^{11}G_{11}=0\ea and
\ba\label{1} \gamma^{00}G_{01}+\gamma^{01}G_{11}=0 .\ea

\subsubsection{The case $Y^\mu (\tau)=0$}
Let us start with considering the particular case, when in
(\ref{tGA}) $Y^\mu (\tau)=0$, i.e. $X^\mu$ depend on $\tau$ and
$\s$ linearly. Then, the Lagrangian density, the induced fields,
the constraints (\ref{0}) and (\ref{1}) respectively, and the
Euler-Lagrange equations for $X^M$ (\ref{em}), can be written as
(the over-dot is used for $d/d\tau$)\ba \label{LRa}
\mathcal{L}^{A}(\tau) = -\frac{T}{2}\sqrt{- \gamma}\left[
\gamma^{00}g_{ab}\dot{Y}^a \dot{Y}^b+ 2\left( \gamma^ {0n}g_{a
\nu}\Lambda_n^\nu - \frac{1}{\sqrt{- \gamma}}\Lambda_1^\nu b_{a
\nu} \right)\dot{Y}^a \right.+
\\ \nl \left. + \gamma^{mn}\Lambda_m^\mu \Lambda_n^\nu g_{\mu \nu}-
\frac{2}{\sqrt{-\gamma}}\Lambda_0^\mu \Lambda_1^\nu b_{\mu \nu}
\right];\ea \ba\label{imtLA} &&G_{00}=g_{ab}\dot{Y}^a\dot{Y}^b +
2\Lambda^\nu_0g_{\nu a}\dot{Y}^a +
\Lambda^\mu_0\Lambda^\nu_0g_{\mu\nu},\\ \nl
&&G_{01}=\Lambda^\nu_1\left(g_{\nu a}\dot{Y}^a +
\Lambda^\mu_0g_{\mu\nu}\right),\h
G_{11}=\Lambda^\mu_1\Lambda^\nu_1g_{\mu\nu};\ea \ba \nl B_{01} =
-\Lambda_1^\mu\left(b_{\mu a}\dot{Y}^a + \Lambda_0^\nu
b_{\mu\nu}\right),\ea
\ba\label{a0}
&&\gamma^{00}g_{ab}\dot{Y}^a\dot{Y}^b +
2\gamma^{00}\Lambda^\nu_0g_{\nu a}\dot{Y}^a +
\left(\gamma^{00}\Lambda^\mu_0\Lambda^\nu_0 -
\gamma^{11}\Lambda^\mu_1\Lambda^\nu_1\right)g_{\mu\nu}=0,
\\
\label{a1} &&\Lambda^\nu_1\left(\gamma^{00}g_{\nu a}\dot{Y}^a +
\gamma^{0n} \Lambda^\mu_n g_{\mu\nu}\right)=0;\ea \ba\label{aem}
&&\gamma^{00}\left(g_{Lb}\ddot{Y}^b +
\Gamma_{L,bc}\dot{Y}^b\dot{Y}^c \right) +
2\gamma^{0n}\Lambda^\mu_n\Gamma_{L,\mu b}\dot{Y}^b +
\gamma^{mn}\Lambda^{\mu}_{m}\Lambda^{\nu}_{n}\Gamma_{L,\mu\nu} \\
\nl &&=-\frac{1}{\sqrt{- \gamma}}\Lambda_1^\nu \left( H_{L \mu
\nu}\Lambda_0^\mu + H_{L a \nu} \dot{Y}^a \right).\ea $
\mathcal{L}^{A}(\tau)$ in (\ref{LRa}) is like a Lagrangian for a
point particle, interacting with the external fields $g_{MN}$,
$b_{a\nu}$ and $b_{\mu\nu}$.

Let us write down the conserved quantities. By definition, the
generalized momenta are \ba \nl P_L\equiv\frac{\p \mathcal{L}}{\p
(\p_0 X^L)} = -T \left(\sqrt{-\gamma}\gamma ^{0n} g_{LN}\p_n X^N -
b_{LN} \p_1 X^N \right). \ea For our ansatz, they take the form:
\ba\nl P_L = -T \left[\sqrt{- \gamma}\left( \gamma^{00} g_{La}
\dot{Y}^a+ \gamma ^{0n}g_{L \nu}\Lambda_n^\nu \right)- b_{L
\nu}\Lambda ^\nu_1 \right].\ea The Lagrangian (\ref{LRa}) does not
depend on the coordinates $X^\mu$. Therefore, the conjugated
momenta $P_\mu$ are conserved \ba\label{cm} \gamma^{00}g_{\mu a}\dot{Y}^a +
\gamma^{0n}\Lambda^{\nu}_{n}g_{\mu\nu} - \frac{1}{\sqrt{-
\gamma}}\Lambda_1 ^\nu b_{\mu\nu} =
-\frac{P_\mu}{T\sqrt{-\gamma}}= constants.\ea
The same result can be obtained by solving the equations of motion
(\ref{aem}) for $L=\lambda$.

From (\ref{a1}) and (\ref{cm}), one obtains the following
compatibility condition \ba\label{cc} \Lambda^{\nu}_{1}P_\nu =
0.\ea This equality may be interpreted as a solution of the
constraint (\ref{a1}), which restricts the number of the
independent parameters in the theory.

With the help of (\ref{cm}), the other constraint, (\ref{a0}), can
be rewritten in the form \ba\label{ec} g_{ab}\dot{Y}^a\dot{Y}^b =
\mathcal{U},\ea where $\mathcal{U}$ is given by \ba\label{sp}
\mathcal{U}=\frac{1}{\gamma^{00}} \left[
\gamma^{mn}\Lambda^{\mu}_{m}\Lambda^{\nu}_{n}g_{\mu\nu} +
\frac{2\Lambda^{\mu}_{0}}{T\sqrt{-\gamma}} \left(P_\mu- T
\Lambda_1^\nu b_{\mu \nu}\right) \right].\ea

Now, let us turn to the equations of motion (\ref{aem}),
corresponding to $L=a$. By using the explicit expressions for
$\Gamma_{a,\mu b}$, $\Gamma_{a,\mu\nu}$, $H_{a\mu\nu}$ and $H_{ab\nu}$,
one obtains \ba\label{fem}
g_{ab}\ddot{Y}^b + \Gamma_{a,bc}\dot{Y}^b\dot{Y}^c =
\frac{1}{2}\p_a \mathcal{U} +
2\p_{[a}\mathcal{A}_{b]}\dot{Y}^b.\ea In (\ref{fem}), an effective
potential $\mathcal{U}$ and an effective gauge field
$\mathcal{A}_a$ appeared. $\mathcal{U}$ is given in (\ref{sp}),
and \ba\label{gf} \mathcal{A}_a= \frac{1}{\gamma
^{00}}\left(\gamma^{0 m}\Lambda_m^\mu g_{a\mu}-
\frac{\Lambda_1^\mu b_{a \mu}}{\sqrt{-\gamma}} \right).\ea

The reduced equations of motion (\ref{fem}) are as for a point
particle moving in the gravitational field $g_{ab}$, in the
potential $\mathcal{U}$ and interacting with the 1-form gauge
field $\mathcal{A}_a$ through its field strength
$\mathcal{F}_{ab}=2\p_{[a}\mathcal{A}_{b]}$. The corresponding
Lagrangian is \ba\nl \mathcal{L}^{A}_{red}(\tau) =
-\frac{T}{2}\sqrt{-\gamma}\gamma^{00}\left( g_{ab}\dot{Y}^a
\dot{Y}^b+ 2\mathcal{A}_a\dot{Y}^a + \mathcal{U}\right) +
\Lambda_0^\mu P_\mu.\ea

Now our task is to find {\it exact} solutions of the {\it
nonlinear} differential equations (\ref{ec}) and (\ref{fem}). It
turns out that for background fields depending on only one
coordinate $x^a$, we can always integrate these equations, and the
solution is \footnote{In this case, the constraint (\ref{ec}) is
first integral for the equation of motion (\ref{fem}).}
\ba\label{ocs}\tau\left(X^a\right)=\tau_0 \pm \int_{X_0^a}^{X^a}d
x \left(\frac{\mathcal{U}}{g_{aa}}\right)^{-1/2}.\ea Otherwise,
supposing the metric $g_{ab}$ is a diagonal one, (\ref{fem}) and
(\ref{ec}) reduce to \ba\label{dem}
&&\frac{d}{d\tau}(g_{aa}\dot{Y}^a) -\frac{1}{2}\left[\p_a g_{aa}(
\dot{Y}^a)^2 +\p_a\mathcal{U}\right]-\frac{1}{2}\sum_{b\ne
a}\left[\p_a g_{bb}(\dot{Y}^b)^2 +
4\p_{[a}\mathcal{A}_{b]}\dot{Y}^b\right] = 0,\\ \label{dec}
&&g_{aa}(\dot{Y}^a)^2+\sum_{b\ne a}
g_{bb}(\dot{Y}^b)^2=\mathcal{U}.\ea With the help of the
constraint (\ref{dec}), we can rewrite the equations of motion
(\ref{dem}) in the form \ba\label{st}
\frac{d}{d\tau}(g_{aa}\dot{Y}^a)^2 - \dot{Y}^a\p_a\left(g_{aa}
\mathcal{U}\right)+ \dot{Y}^a\sum_{b\ne a}
\left[\p_a\left(\frac{g_{aa}}{g_{bb}}\right) (g_{bb}\dot{Y}^b)^2 -
4g_{aa}\p_{[a}\mathcal{A}_{b]} \dot{Y}^b\right] = 0.\ea

To find solutions of the above equations without choosing
particular background, we can fix all coordinates $Y^a$ except
one. Then the $exact$ string solution of the equations of motion
and constraints is given again by the same expression (\ref{ocs})
for $\tau\left(X^a\right)$.

To find solutions depending on more than one coordinate, we have
to impose further conditions on the background fields. Let us
first consider the simpler case, when the last two terms in
(\ref{st}) are not present. This may happen, when \ba\label{sc}
\p_a\left(\frac{g_{aa}}{g_{bb}}\right) = 0,\h
\mathcal{A}_{a}=0.\ea Then, the first integrals of (\ref{st}) are
\ba\label{fias} \left(g_{aa}\dot{Y}^a\right)^2 =D_a(Y^{b\ne
a})+g_{aa}\mathcal{U},\ea where $D_a$ are arbitrary functions of
their arguments. These solutions must be compatible with the
constraint (\ref{dec}), which leads to the condition
\ba\nl\sum_a\frac{D_a}{g_{aa}}=(1-n_a)\mathcal{U},\ea where $n_a$
is the number of the coordinates $Y^a$. From here, one can express
one of the functions $D_a$ through the others. To this end, we
split the index $a$ in such a way that $Y^r$ is one of the
coordinates $Y^a$, and $Y^{\alpha}$ are the others. Then \ba\nl
D_r=-g_{rr}\left(n_{\alpha}\mathcal{U}
+\sum_{\alpha}\frac{D_{\alpha}}{g_{\alpha\alpha}}\right),\ea and
by using this, one rewrites the first integrals (\ref{fias}) as
\ba\label{fiasf} \left(g_{rr}\dot{Y}^r\right)^2 = g_{rr}
\left[(1-n_{\alpha})\mathcal{U} - \sum_{\alpha}
\frac{D_{\alpha}}{g_{\alpha\alpha}}\right]\ge 0,\h
\left(g_{\alpha\alpha}\dot{Y}^\alpha\right)^2 =D_\alpha(Y^{a\ne
\alpha})+g_{\alpha\alpha}\mathcal{U}\ge 0, \ea where $n_{\alpha}$
is the number of the coordinates $Y^\alpha$. Thus, the constraint
(\ref{dec}) is satisfied identically.

Now we turn to the general case, when all terms in the equations
of motion (\ref{st}) are present. The aim is to find conditions,
which will allow us to reduce the order of the equations of motion
by one. An example of such {\it sufficient} conditions, is given
below : \ba\nl &&\mathcal{A}_a\equiv\left(
\mathcal{A}_r,\mathcal{A}_{\alpha} \right)= \left(
\mathcal{A}_r,\p_{\alpha}f \right),\h
\p_{\alpha}\left(\frac{g_{\alpha\alpha}}{g_{aa}}\right)=0,\\ \nl
&&\p_{\alpha}\left(g_{rr}\dot{Y}^r\right)^2 = 0,\h
\p_{r}\left(g_{\alpha\alpha}\dot{Y}^{\alpha}\right)^2 = 0.\ea By
using the restrictions given above, one obtains the following
first integrals of the equations (\ref{st}), compatible with the
constraint (\ref{dec}) \ba\label{fir}
\left(g_{rr}\dot{Y}^{r}\right)^2 &=&
g_{rr}\left[\left(1-n_{\alpha}\right) \mathcal{U} -
\sum_{\alpha}\frac{D_{\alpha}}{g_{\alpha\alpha}} -
2n_{\alpha}\left(\mathcal{A}_{r}-\p_r f\right)\dot{Y}^{r}\right]=
E_r\left(Y^r\right)\ge 0,\\ \label{fia}
\left(g_{\alpha\alpha}\dot{Y}^{\alpha}\right)^2 &=& D_{\alpha}
\left(Y^{a\ne\alpha}\right) +
g_{\alpha\alpha}\left[\mathcal{U}+2\left( \mathcal{A}_{r}-\p_r
f\right)\dot{Y}^r\right]= E_{\alpha}\left(Y^{\beta}\right)\ge 0,
\ea where $D_{\alpha}$, $E_{\alpha}$ and $E_r$ are arbitrary
functions of their arguments.

Further progress is possible, when working with particular
background configurations, allowing for separation of the
variables in (\ref{fiasf}), or in (\ref{fir}) and (\ref{fia}).

Our results obtained so far are not applicable to tensionless
(null) strings, because the action (\ref{pa}) is proportional to
the string tension $T$. The parametrization of $\gamma^{mn}$,
which is appropriate for considering the zero tension limit $T\to
0$, is the following \cite{ILST93, HLU94}: \ba\label{tl}
\gamma^{00}=-1,\h \gamma^{01}=\lambda^1,\h
\gamma^{11}=(2\lambda^0T)^2 - (\lambda^1)^2, \h \det(\gamma^{mn})=
-(2\lambda^0T)^2.\ea Here $\lambda^n$ are the Lagrange
multipliers, whose equations of motion generate the {\it
independent} constraints. In these notations, the constraints
(\ref{a0}) and (\ref{a1}), the equations of motion (\ref{aem}),
and the conserved momenta (\ref{cm}) take the form \ba\nl
&&g_{ab}\dot{Y}^a\dot{Y}^b + 2\Lambda^\nu_0g_{\nu a}\dot{Y}^a +
\left\{\Lambda^\mu_0\Lambda^\nu_0
+\left[(2\lambda^0T)^2-(\lambda^1)^2\right]
\Lambda^\mu_1\Lambda^\nu_1\right\} g_{\mu\nu}=0,\\ \nl
&&\Lambda^\nu_1\left[g_{\nu a}\dot{Y}^a +
\left(\Lambda^\mu_0-\lambda^1\Lambda^\mu_1\right)
g_{\mu\nu}\right]=0;\ea \ba\nl &&g_{Lb}\ddot{Y}^b +
\Gamma_{L,bc}\dot{Y}^b\dot{Y}^c +
2\left(\Lambda^\mu_0-\lambda^1\Lambda^\mu_1\right)\Gamma_{L,\mu
b}\dot{Y}^b\\ \nl
&&+\left[\left(\Lambda^\mu_0-\lambda^1\Lambda^\mu_1\right)
\left(\Lambda^\nu_0-\lambda^1\Lambda^\nu_1\right)-(2\lambda^0T)^2
\Lambda^\mu_1\Lambda^\nu_1\right]\Gamma_{L,\mu\nu}= 2 \lambda^0 T
\Lambda_1^\nu \left(H_{L \nu b}\dot{Y}^b + \Lambda_0^\mu H_{L\mu
\nu}\right);\ea \ba\nl g_{\mu a}\dot{Y}^a +
\left(\Lambda^\nu_0-\lambda^1\Lambda^\nu_1\right)g_{\mu\nu} +
2\lambda^0T\Lambda^\nu_1b_{\mu\nu}= 2\lambda^0P_\mu.\ea The
reduced equations of motion and constraint (\ref{fem}) and
(\ref{ec}) have the same form, but now, the effective potential
(\ref{sp}) and the effective gauge field (\ref{gf}) are given by
\ba\nl &&\mathcal{U}^{\lambda}=
\left[\left(\Lambda^\mu_0-\lambda^1\Lambda^\mu_1\right)
\left(\Lambda^\nu_0-\lambda^1\Lambda^\nu_1\right)-(2\lambda^0T)^2
\Lambda^\mu_1\Lambda^\nu_1\right]g_{\mu\nu}
-4\lambda^0\Lambda_0^\mu\left( P_\mu - T\Lambda_1^\nu
b_{\mu\nu}\right),\\ \nl &&\mathcal{A}^{\lambda}_a
=\left(\Lambda^\mu_0-\lambda^1\Lambda^\mu_1\right) g_{a\mu} + 2
\lambda^0 T\Lambda_1^\mu b_{a \mu}.\ea

If one sets $\lambda^1=0$ and $2\lambda^0T=1$, the results in {\it
conformal gauge}  are obtained, as it should be. If one puts $T=0$
in the above formulas, they will describe {\it tensionless}
strings.

\subsubsection{The case $Y^\mu (\tau)\ne 0$}
By using the ansatz (\ref{tGA}), one obtains that the Lagrangian
density, the induced fields, the constraints (\ref{0}) and
(\ref{1}) respectively, and the Euler-Lagrange equations for $X^M$
(\ref{em}) are given by \ba\nl \mathcal{L}^{GA}(\tau) =
-\frac{T}{2}\sqrt{-\gamma}\left[\gamma^{00}g_{MN} \dot{Y}^M
\dot{Y}^N + 2 \left(\gamma^{0n}\Lambda_n^\nu g_{M \nu}-
\frac{\Lambda_1^\nu b_{M \nu}}{\sqrt{- \gamma}} \right)\dot{Y}^M +
\right.\\ \nl +\left. \gamma^{mn}\Lambda_m^\mu \Lambda_n^\nu
g_{\mu \nu}-\frac{2\Lambda _0^\mu\Lambda_1^\nu
b_{\mu\nu}}{\sqrt{-\gamma}} \right];\ea \ba \label{imga}
&&G_{00}=g_{MN}\dot{Y}^M\dot{Y}^N + 2\Lambda^\nu_0g_{\nu
N}\dot{Y}^N + \Lambda^\mu_0\Lambda^\nu_0g_{\mu\nu},\\ \nl
&&G_{01}=\Lambda^\nu_1\left(g_{\nu N}\dot{Y}^N +
\Lambda^\mu_0g_{\mu\nu}\right),\h
G_{11}=\Lambda^\mu_1\Lambda^\nu_1g_{\mu\nu};\ea \ba\nl
B_{01} = -\Lambda_1^\mu\left(b_{\mu N}\dot{Y}^N+
\Lambda_0^\nu b_{\mu\nu}\right),\ea
\ba\label{mga0} &&\gamma^{00}g_{MN}\dot{Y}^M\dot{Y}^N +
2\gamma^{00}\Lambda^\nu_0g_{\nu N}\dot{Y}^N +
\left(\gamma^{00}\Lambda^\mu_0\Lambda^\nu_0 -
\gamma^{11}\Lambda^\mu_1\Lambda^\nu_1\right)g_{\mu\nu}=0,\\
\label{mga1} &&\Lambda^\nu_1\left(\gamma^{00}g_{\nu N}\dot{Y}^N +
\gamma^{0n} \Lambda^\mu_n g_{\mu\nu}\right)=0;\ea \ba\label{mgaem}
\gamma^{00}\left(g_{LN}\ddot{Y}^N +
\Gamma_{L,MN}\dot{Y}^M\dot{Y}^N \right) +
2\gamma^{0n}\Lambda^\mu_n\Gamma_{L,\mu N}\dot{Y}^N +
\gamma^{mn}\Lambda^{\mu}_{m}\Lambda^{\nu}_{n}\Gamma_{L,\mu\nu}= \\
\nl = -\frac{1}{\sqrt{- \gamma}}\Lambda_1 ^\nu\left(H_{LM \nu}
\dot{Y}^M + \Lambda_0^\mu H_{L \mu\nu}\right) .\ea The conserved
momenta $P_\mu$ can be found as before, and now they are
\ba\label{mgcm} \gamma^{00}g_{\mu N}\dot{Y}^N +
\gamma^{0n}\Lambda^{\nu}_{n}g_{\mu\nu}- \frac{\Lambda_1^\nu
b_{\mu\nu}}{\sqrt{- \gamma}} = -\frac{P_\mu}{T\sqrt{-\gamma}}=
constants.\ea The compatibility condition following from the
constraint (\ref{mga1}) and from (\ref{mgcm}) coincides with the
previous one (\ref{cc}). With the help of (\ref{mgcm}), the
equations of motion (\ref{mgaem}) corresponding to $L=a$ and the
other constraint (\ref{mga0}), can be rewritten in the form
\ba\label{mgmem} &&g_{aN}\ddot{Y}^N +
\Gamma_{a,MN}\dot{Y}^M\dot{Y}^N = \frac{1}{2}\p_a \mathcal{U} +
2\p_{[a}\mathcal{A}_{N]}\dot{Y}^N,\\ \label{mgec}
&&g_{MN}\dot{Y}^M\dot{Y}^N = \mathcal{U},\ea where $\mathcal{U}$
is given by (\ref{sp}) and \ba\label{mggm} \mathcal{A}_N=
\frac{1}{\gamma ^{00}}\left(\gamma^{0 m}\Lambda_m^\mu g_{N\mu}-
\frac{\Lambda_1^\mu b_{N \mu}}{\sqrt{-\gamma}} \right) \ea
coincides with (\ref{gf}) for $N=a$.

Now we are going to eliminate the variables $\dot{Y}^\mu$ from
(\ref{mgmem}) and (\ref{mgec}). To this end, we express
$\dot{Y}^\mu$ through $\dot{Y}^a$ from the conservation laws
(\ref{mgcm}): \ba\label{muc} \dot{Y}^\mu =
-\frac{\gamma^{0n}}{\gamma^{00}}\Lambda^\mu_n -
\left(g^{-1}\right)^{\mu\nu}\left[g_{\nu a}\dot{Y}^a +
\frac{1}{\gamma^{00}T\sqrt{-\gamma}}\left( P_\nu - T\Lambda_1^\rho
b_{\nu \rho} \right)\right].\ea

After using (\ref{muc}) and (\ref{cc}), the equations of motion
(\ref{mgmem}) and the constraint (\ref{mgec}) acquire the form
\ba\label{mgemf} &&h_{ab}\ddot{Y}^b +
\Gamma^{\bf{h}}_{a,bc}\dot{Y}^b\dot{Y}^c = \frac{1}{2}\p_a
\mathcal{U}^{\bf{h}} +
2\p_{[a}\mathcal{A}^{\bf{h}}_{b]}\dot{Y}^b,\\ \label{mgecf}
&&h_{ab}\dot{Y}^a\dot{Y}^b = \mathcal{U}^{\bf{h}},\ea where a new,
effective metric appeared \ba\nl h_{ab} = g_{ab} -
g_{a\mu}(g^{-1})^{\mu\nu}g_{\nu b}.\ea $\Gamma^{\bf{h}}_{a,bc}$ is
the symmetric connection corresponding to this metric \ba\nl
\Gamma^{\bf{h}}_{a,bc}=\frac{1}{2}\left(\p_bh_{ca}
+\p_ch_{ba}-\p_ah_{bc}\right).\ea The new effective scalar and
gauge potentials, expressed through the background fields, are as
follows \ba\nl
&&\mathcal{U}^{\bf{h}}=\frac{1}{\gamma\left(\gamma^{00}\right)^2}
\left[ \Lambda^{\mu}_{1}\Lambda^{\nu}_{1}g_{\mu\nu}+ \frac{1}{T^2}
\left( P_\mu - T\Lambda_1^\rho b_{\mu \rho}\right)
(g^{-1})^{\mu\nu} \left( P_\nu - T\Lambda_1^\lambda
b_{\nu\lambda}\right)\right],
\\ \nl &&\mathcal{A}^{\bf{h}}_{a}= -
\frac{1}{\gamma^{00}T\sqrt{-\gamma}} \left[g_{a\mu}
(g^{-1})^{\mu\nu} \left(P_\nu - T\Lambda _1 ^ \rho b_{\nu\rho}
\right)+ T \Lambda_1 ^\rho b_{a \rho} \right] .\ea We point out
the qualitatively different behavior of the potentials
$\mathcal{U}^{\bf{h}}$ and $\mathcal{A}_a^{\bf{h}}$, compared to
$\mathcal{U}$ and $\mathcal{A}_a$, due to the appearance of the
inverse metric $(g^{-1})^{\mu\nu}$. The corresponding Lagrangian is
\ba\nl \mathcal{L}^{GA}_{red}(\tau) =
-\frac{T}{2}\sqrt{-\gamma}\gamma^{00}\left(h_{ab} \dot{Y}^a
\dot{Y}^b + 2\mathcal{A}_a^{\bf{h}}\dot{Y}^a + \mathcal{U}^{\bf{h}}
\right) + \frac{d}{d\tau}P_\mu\left(Y^\mu+\Lambda_0^\mu \tau\right).\ea
Since the equations (\ref{fem}), (\ref{ec}) and (\ref{mgemf}),
(\ref{mgecf}) have the same form, for obtaining exact string
solutions, we  can proceed as before and use the previously
derived formulas after the replacements
$(g,\Gamma,\mathcal{U},\mathcal{A})$ $\to$ $(h,\Gamma^{\bf{h}},
\mathcal{U}^{\bf{h}},\mathcal{A}^{\bf{h}})$. In particular, the
solution depending on one of the coordinates $X^a$ will be
\ba\label{mgocs}\tau\left(X^a\right)=\tau_0 \pm
\int_{X_0^a}^{X^a}d x
\left(\frac{\mathcal{U}^{\bf{h}}}{h_{aa}}\right)^{-1/2}.\ea In
this case by integrating (\ref{muc}), and replacing the solution
for $Y^\mu$ in the ansatz (\ref{tGA}), one obtains the solution
for the string coordinates $X^\mu$: \ba \label{X} &&X^\mu(X^a,
\sigma) = X_0^\mu + \Lambda_1 ^\mu \left[ \sigma -
\frac{\gamma^{01}}{\gamma^{00}}\tau\left(X^a\right) \right] -\\
\nl && - \int_ {X_0^a}^ {X^a} (g^{-1})^{\mu\nu} \left[g_{\nu a}\pm
\frac{\left(P_\nu - T\Lambda_1^\rho b_{\nu\rho}\right)}
{\gamma^{00}T \sqrt{- \gamma}} \left(\frac{\mathcal
{U}^{\rm{h}}}{h_{aa}} \right)^{-1/2} \right] d x .\ea

To be able to take the tensionless limit $T\to 0$ in the above
formulas, we have to use the $\lambda$-parametrization (\ref{tl})
of $\gamma^{mn}$. The quantities that appear in the reduced
equations of motion and constraint (\ref{mgemf}) and
(\ref{mgecf}), which depend on this parametrization, are
$\mathcal{U}^{\bf{h}}$ and $\mathcal{A}^{\bf{h}}_{a}$. Now, they
are given by \ba\nl && \mathcal{U}^{h,\lambda}= - (2\lambda^0)^2
\left[T^2 \Lambda_1^\mu \Lambda_1 ^\nu g_{\mu\nu} + \left(P_\mu -
T\Lambda_1^\rho b_{\mu\rho} \right) (g^{-1})^{\mu\nu}
\left(P_\nu-T\Lambda_1^\lambda b_{\nu\lambda}\right)\right], \\
\nl && \mathcal{A}_{a}^{h,\lambda} = 2\lambda^0
\left[g_{a\mu}(g^{-1})^{\mu\lambda} (P_\lambda - T\Lambda_1^\rho
b_{\lambda\rho}) + T\Lambda_1^\rho b_{a\rho} \right]. \ea If one
sets $\lambda^1=0$ and $2\lambda^0T=1$, the {\it conformal gauge}
results are obtained. If one puts $T=0$ in the above equalities,
they will correspond to {\it tensionless} strings.

\subsection{$\s$-dependent dynamics}
In this subsection, we will use the ansatz (\ref{sGA}) for the
string embedding coordinates. Of course, the conditions (\ref{ob})
on the background fields are also fulfilled.

Our {\it independent} constraints, with which we will work in this
subsection, are given by \ba\label{0s}
\gamma^{00}G_{00}-\gamma^{11}G_{11}=0,\ea and \ba\label{1s}
\gamma^{01}G_{00}+\gamma^{11}G_{01}=0 .\ea

\subsubsection{The case $Z^\mu (\s)= 0$}
Taking into account the
conditions (\ref{ob}), one obtains the following reduced
Lagrangian density, arising from the action (\ref{pa}) (the prime
is used for $d/d\sigma$)\ba\label{La}\mathcal{L}^{A}(\sigma) =
-\frac{T}{2}\sqrt{- \gamma}\left[ \gamma^{11}g_{ab}Z'^aZ'^b+
2\left( \gamma^ {1m}\Lambda_m^\mu g_{\mu a} - \frac{1}{\sqrt{-
\gamma}}\Lambda_0^\mu b_{\mu a} \right)Z'^a \right.+ \\ \nl \left.
+ \gamma^{mn}\Lambda_m^\mu \Lambda_n^\nu g_{\mu \nu}-
\frac{2}{\sqrt{-\gamma}}\Lambda_0^\mu \Lambda_1^\nu b_{\mu \nu}
\right],\ea where the fields induced on the string worldsheet are
given by \ba\nl &&G_{00}=\Lambda^\mu_0\Lambda^\nu_0 g_{\mu\nu},\h
G_{01}=\Lambda^\mu_0\left(g_{\mu a}Z'^a + \Lambda^\nu_1
g_{\mu\nu}\right),\\ \label{imsLA} &&G_{11}=g_{ab}Z'^a Z'^b +
2\Lambda^\mu_1 g_{\mu a}Z'^a + \Lambda^\mu_1\Lambda^\nu_1
g_{\mu\nu} ;\ea \ba\nl B_{01}=\Lambda^\mu_0\left(b_{\mu a}Z'^a +
\Lambda^\nu_1 b_{\mu\nu}\right);\ea

The constraints (\ref{0s}) and (\ref{1s}) respectively, and the
equations of motion for $X^M$ (\ref{em}), can be written as
\ba\label{a0s} &&\gamma^{11}\left(g_{ab}Z'^aZ'^b +
2\Lambda^\mu_1g_{\mu b}Z'^b\right) -
\left(\gamma^{00}\Lambda^\mu_0\Lambda^\nu_0 -
\gamma^{11}\Lambda^\mu_1\Lambda^\nu_1\right)g_{\mu\nu}=0,\\
\label{a1s} &&\Lambda^\mu_0\left(\gamma^{11}g_{\mu a}Z'^a +
\gamma^{1n} \Lambda^\nu_n g_{\mu\nu}\right)=0;\ea \ba\label{aems}
&&\gamma^{11}\left(g_{Lb}Z''^b + \Gamma_{L,bc}Z'^bZ'^c \right) +
2\gamma^{1m}\Lambda^\mu_m\Gamma_{L,\mu b}Z'^b +
\gamma^{mn}\Lambda^{\mu}_{m}\Lambda^{\nu}_{n}\Gamma_{L,\mu\nu} \\
\nl &&=-\frac{1}{\sqrt{- \gamma}}\Lambda_0^\mu \left(H_{L\mu
a}Z'^a + \Lambda_1^\nu H_{L \mu \nu}\right).\ea

Let us write down the conserved quantities. By definition, the
generalized momenta are \ba \nl P_L\equiv\frac{\p
\mathcal{L}^P}{\p (\p_0 X^L)} = -T \left(\sqrt{-\gamma}\gamma
^{0n} g_{LN}\p_n X^N - b_{LN} \p_1 X^N \right). \ea For our case,
they take the form: \ba\nl P_L = -T \sqrt{-
\gamma}\left[ \left(\gamma^{01} g_{Lb} - \frac{1}{\sqrt{-
\gamma}}b_{Lb}\right) Z'^b + \gamma ^{0n}\Lambda_n^\nu g_{L \nu}-
\frac{1}{\sqrt{- \gamma}}\Lambda ^\nu_1 b_{L \nu}\right].\ea The
Lagrangian (\ref{La}) does not depend on the coordinates $X^\mu$.
Therefore, the conjugated momenta $P_\mu$ do not depend on the
proper time $\tau$ \footnote{Actually, all momenta $P_M$ do not
depend on $\tau$, because there is no such dependence in
(\ref{La}).} \ba\label{cmla} P_\mu(\s) = -T \sqrt{- \gamma}
\left[ \left(\gamma^{01} g_{\mu b} - \frac{1}{\sqrt{-
\gamma}}b_{\mu b}\right) Z'^b+\gamma ^{0n}\Lambda_n^\nu g_{\mu
\nu}- \frac{1}{\sqrt{- \gamma}}\Lambda ^\nu_1 b_{\mu
\nu}\right],\hspace{.1cm} \p_0 P_\mu=0 .\ea

In order for our ansatz to be  consistent
with the action (\ref{pa}), the following conditions must be
fulfilled \ba\label{emmu} \p_1\mathcal{P}_\mu\equiv
\frac{\p\mathcal{P}_\mu}{\p\s}=0,\ea where \ba\nl &&\mathcal{P}_M
\equiv\frac{\p \mathcal{L}^P}{\p (\p_1 X^M)} =-T\left(\sqrt{-
\gamma}\gamma ^{1n}g_{MN}\p_n X^N + b_{MN}\p_0 X^N\right)\\
\label{pM}&&=-T \sqrt{- \gamma}\left[ \gamma^{11} g_{Mb}Z'^b +
\gamma ^{1n}\Lambda_n^\nu g_{M\nu}+ \frac{1}{\sqrt{-
\gamma}}\Lambda ^\nu_0 b_{M\nu}\right].\ea This is because the
equations of motion (\ref{em}) can be rewritten as \ba\nl \frac{\p
P_M }{\p\tau} + \frac{\p \mathcal{P}_M}{\p\sigma} - \frac{\p
\mathcal{L}^P}{\p x^M}=0.\ea Hence, for $M=\mu$, these equations take
the form (\ref{emmu}). Therefore, $\mathcal{P}_\mu$
are constants of the motion \ba\label{cq}
\gamma^{11}g_{\mu a}Z'^a + \gamma^{1n}\Lambda^{\nu}_{n}g_{\mu\nu}
+ \frac{1}{\sqrt{- \gamma}}\Lambda_0 ^\nu b_{\mu\nu} =
-\frac{\mathcal{P}_\mu}{T\sqrt{-\gamma}} = constants.\ea

From (\ref{a1s}) and (\ref{cq}), one obtains the following
compatibility condition \ba\label{ccs}
\Lambda^{\nu}_{0}\mathcal{P}_\nu = 0.\ea This equality may be
interpreted as a solution of the constraint (\ref{a1s}), which
restricts the number of the independent parameters in the theory.

With the help of (\ref{cq}), the other constraint, (\ref{a0s}), can
be rewritten in the form \ba\label{ecs} g_{ab}Z'^aZ'^b =
\mathcal{U}^{s},\ea where $\mathcal{U}^{s}$ is given by
\ba\label{sps} \mathcal{U}^{s}=\frac{1}{\gamma^{11}} \left[
\gamma^{mn}\Lambda^{\mu}_{m}\Lambda^{\nu}_{n}g_{\mu\nu} +
\frac{2\Lambda^{\mu}_{1}}{T\sqrt{-\gamma}} \left(\mathcal{P}_\mu+
T \Lambda_0^\nu b_{\mu \nu}\right) \right].\ea

Now, let us turn to the equations of motion (\ref{aem}),
corresponding to $L=a$. In view of the conditions (\ref{ob}),
\ba\nl &&\Gamma_{a,\mu
b}=-\frac{1}{2}\left(\p_ag_{b\mu}-\p_bg_{a\mu}\right)
=-\p_{[a}g_{b]\mu},\h
\Gamma_{a,\mu\nu}=-\frac{1}{2}\p_ag_{\mu\nu},\\ \nl &&
H_{a\mu\nu}= \p_a b_{\mu\nu}; \h H_{ab\nu}=\p_ab_{b\nu} - \p_b
b_{a \nu}= 2\p_{[a}b_{b]\nu}.\ea By using this, one obtains
\ba\label{fems} g_{ab}Z''^b + \Gamma_{a,bc}Z'^bZ'^c =
\frac{1}{2}\p_a \mathcal{U}^{s} +
2\p_{[a}\mathcal{A}^{s}_{b]}Z'^b.\ea In (\ref{fems}), an effective
scalar potential $\mathcal{U}^{s}$ and an effective 1-form gauge
field $\mathcal{A}^{s}_a$ appeared. $\mathcal{U}^{s}$ is given in
(\ref{sps}) (and is the same as in the effective constraint
(\ref{ecs})), and \ba\label{gfs} \mathcal{A}^{s}_a= \frac{1}{\gamma
^{11}}\left(\gamma^{1 m}\Lambda_m^\mu g_{a\mu}+
\frac{1}{\sqrt{-\gamma}}\Lambda_0^\mu b_{a \mu} \right).\ea The
corresponding Lagrangian is \ba\nl \mathcal{L}_{red}^{A}(\sigma) =
-\frac{T}{2}\sqrt{- \gamma}\gamma^{11}\left(g_{ab}Z'^aZ'^b+
2\mathcal{A}^{s}_a Z'^a + \mathcal{U}^{s}\right) +
\Lambda_1^{\mu}\mathcal{P}_{\mu}.\ea

Now our task is to find {\it exact} solutions of the {\it
nonlinear} differential equations (\ref{ecs}) and (\ref{fems}).

If the background seen by the string depends on only one
coordinate $x^a$, the general solution for the string embedding
coordinate $X^a(\tau,\s)=Z^a(\s)$ is given by
\ba\nl\sigma\left(X^a\right)=\sigma_0 + \int_{X_0^a}^{X^a}
\left(\frac{\mathcal{U}^{s}}{g_{aa}}\right)^{-1/2}dx .\ea When the
background felt by the string depends on more than one coordinate
$x^a$, the first integrals of the equations of motion for
$Z^a(\s)=(Z^r, Z^\alpha)$, which also solve the constraint
(\ref{ecs}), are \ba\nl &&\left(g_{rr}Z'^r\right)^2 = g_{rr}
\left[(1-n_\alpha)\mathcal{U}^{s} -2n_\alpha\left(\mathcal{A}^{s}_r-\p_r
f\right)Z'^r -\sum_{\alpha}\frac{D_{\alpha}
\left(Z^{a\ne\alpha}\right)} {g_{\alpha\alpha}}\right]=
F_r\left(Z^r\right)\ge 0,\\ \nl
&&\left(g_{\alpha\alpha}Z'^\alpha\right)^2 =D_{\alpha}
\left(Z^{a\ne\alpha}\right) + g_{\alpha\alpha}\left[\mathcal{U}^{s}
+2\left(\mathcal{A}^{s}_r-\p_r f\right)Z'^r\right]= F_{\alpha}
\left(Z^{\beta}\right)\ge 0,\ea where $Z^r$ is one of the
coordinates $Z^a$, $Z^\alpha$ are the remaining ones, $n_\alpha$
is the number of $Z^\alpha$, and $D_{\alpha}$, $F_{a}$ are
arbitrary functions of their arguments. The above expressions are
valid, if the $g_{ab}$ part of the metric is diagonal one, and the
following integrability conditions hold
\ba\nl &&\mathcal{A}^{s}_a \equiv(\mathcal{A}^{s}_r,\mathcal{A}^{s}_\alpha)=
(\mathcal{A}^{s}_r,\p_\alpha f),\h
\p_\alpha\left(\frac{g_{\alpha\alpha}}{g_{aa}}\right)=0,
\\ \nl &&\p_\alpha\left(g_{rr}Z'^r\right)^2=0,\h
\p_r\left(g_{\alpha\alpha}Z'^\alpha\right)^2=0. \ea

In the parametrization (\ref{tl}) of $\gamma^{mn}$, the action (\ref{pa})
becomes
\ba\nl S_\lambda =\int d^2\xi\Bigl\{\frac{1}{4\lambda^0}\Bigl[
G_{00}-2\lambda^{1}G_{01}+\left(\lambda^{1}\right)^2 G_{11}
-\left(2\lambda^0T\right)^2 G_{11}\Bigr] + TB_{01}\Bigr\}.\ea

In these notations, the constraints (\ref{a0s}) and (\ref{a1s}), the
equations of motion (\ref{aems}), and the conserved quantities
(\ref{cmla}), (\ref{cq}) take the form \ba\nl &&g_{ab}Z'^aZ'^b +
2\Lambda_{1}^{\mu}g_{\mu b}Z'^b +
\left[\frac{\Lambda^\mu_0\Lambda^\nu_0}{(2\lambda^0T)^2 -
(\lambda^1)^2}+\Lambda^\mu_1\Lambda^\nu_1\right]g_{\mu\nu}=0,\\
\nl &&\Lambda^\mu_0\left\{g_{\mu a}Z'^a +
\left[\frac{\lambda^1\Lambda^\nu_0}{(2\lambda^0T)^2 -
(\lambda^1)^2}+\Lambda^\nu_1\right]g_{\mu\nu}\right\}=0;\ea \ba\nl
&&g_{Lb}Z''^b + \Gamma_{L,bc}Z'^bZ'^c +
2\left[\frac{\lambda^1\Lambda^\mu_0}{(2\lambda^0T)^2 -
(\lambda^1)^2}+\Lambda^\mu_1\right]\Gamma_{L,\mu b}Z'^b\\ \nl &&+
\left[\frac{\Lambda^\mu_0\left(2\lambda^1\Lambda^\nu_1 -
\Lambda^\nu_0 \right)}{(2\lambda^0T)^2 - (\lambda^1)^2}
+\Lambda^\mu_1\Lambda^\nu_1\right]\Gamma_{L,\mu\nu}
=-\frac{2\lambda^0 T}{(2\lambda^0T)^2 -
(\lambda^1)^2}\Lambda_0^\mu \left(H_{L\mu a}Z'^a + \Lambda_1^\nu
H_{L \mu \nu}\right).\ea \ba\nl &&P_\mu(\s)
=\frac{1}{2\lambda^0}\left[\left(-\lambda^1g_{\mu a} +
2\lambda^0Tb_{\mu a}\right)Z'^a +
\left(\Lambda^{\nu}_0-\lambda^1\Lambda^{\nu}_1\right)g_{\mu\nu} +
2\lambda^0T\Lambda^{\nu}_1b_{\mu\nu}\right],\\ \nl
&&\mathcal{P}_{\mu} =-\frac{1}{2\lambda^0}
\left\{\left[(2\lambda^0T)^2-(\lambda^1)^2\right]\left(g_{\mu a}
Z'^{a}+ \Lambda^{\nu}_1 g_{\mu\nu}\right)+ \Lambda^{\nu}_0
\left(\lambda^1 g_{\mu\nu} + 2\lambda^0 T
b_{\mu\nu}\right)\right\}.\ea

The reduced equations of motion and constraint (\ref{fems}) and
(\ref{ecs}) have the same form, but now, the effective potential
(\ref{sps}) and the effective gauge field (\ref{gfs}) are given by
\ba\nl &&\mathcal{U}^{s,\lambda} =
\left[\frac{\Lambda^\mu_0\left(2\lambda^1\Lambda^\nu_1 -
\Lambda^\nu_0 \right)}{(2\lambda^0T)^2 - (\lambda^1)^2}
+\Lambda^\mu_1\Lambda^\nu_1\right]g_{\mu\nu} +
\frac{4\lambda^0}{(2\lambda^0T)^2 -
(\lambda^1)^2}\Lambda^\mu_1\left( \mathcal{P}_\mu + T\Lambda^\nu_0
b_{\mu\nu}\right),\\ \nl &&\mathcal{A}^{s,\lambda}_a =
\left[\frac{\lambda^1\Lambda^\nu_0}{(2\lambda^0T)^2 -
(\lambda^1)^2}+\Lambda^\nu_1\right]g_{a\nu} + \frac{2\lambda^0
T}{(2\lambda^0T)^2 - (\lambda^1)^2}\Lambda_0^\mu b_{a\mu}.\ea

If one sets $\lambda^1=0$ and $2\lambda^0T=1$, this will
correspond to {\it conformal gauge}, as it should be. If one puts
$T=0$ in the above formulas, they will describe {\it tensionless}
strings.

\subsubsection{The case $Z^\mu (\s)\ne 0$}
Taking into account the ansatz (\ref{sGA}), one obtains that the
induced fields $G_{mn}$ and $B_{mn}$, the Lagrangian density, the
constraints (\ref{0s}) and (\ref{1s}) respectively, and the
Euler-Lagrange equations for $X^M$ (\ref{em}) are given by \ba
\label{imgas} &&G_{00}=\Lambda^\mu_0\Lambda^\nu_0 g_{\mu\nu},\h
G_{01}=\Lambda^\mu_0\left(g_{\mu N}Z'^N + \Lambda^\nu_1
g_{\mu\nu}\right),\\ \nl &&G_{11}=g_{MN}Z'^M Z'^N + 2\Lambda^\mu_1
g_{\mu N}Z'^N + \Lambda^\mu_1\Lambda^\nu_1 g_{\mu\nu} ;\ea \ba\nl
B_{01}=\Lambda^\mu_0\left(b_{\mu N}Z'^N + \Lambda^\nu_1
b_{\mu\nu}\right);\ea \ba\nl\mathcal{L}^{GA}(\sigma) =
-\frac{T}{2}\sqrt{-\gamma}\left[\gamma^{11}g_{MN}Z'^MZ'^N + 2
\left(\gamma^{1m}\Lambda_m^\mu g_{\mu N}- \frac{\Lambda_0^\mu b_{
\mu N}}{\sqrt{- \gamma}} \right)Z'^N + \right.\\ \nl +\left.
\gamma^{mn}\Lambda_m^\mu \Lambda_n^\nu g_{\mu \nu}- \frac{2
\Lambda_0^\mu\Lambda_1^\nu b_{\mu\nu}}{\sqrt{-\gamma}} \right];\ea
\ba\label{mga0s} &&\gamma^{11}g_{MN}Z'^M Z'^N +
2\gamma^{11}\Lambda^\mu_1g_{\mu N}Z'^N -
\left(\gamma^{00}\Lambda^\mu_0\Lambda^\nu_0 -
\gamma^{11}\Lambda^\mu_1\Lambda^\nu_1\right)g_{\mu\nu}=0,\\
\label{mga1s} &&\Lambda^\mu_0\left(\gamma^{11}g_{\mu N}Z'^N +
\gamma^{1n} \Lambda^\nu_n g_{\mu\nu}\right)=0;\ea \ba\label{mgaems}
\gamma^{11}\left(g_{LN}Z''^N + \Gamma_{L,MN}Z'^M Z'^N \right) +
2\gamma^{1m}\Lambda^\mu_m\Gamma_{L,\mu N}Z'^N +
\gamma^{mn}\Lambda^{\mu}_{m}\Lambda^{\nu}_{n}\Gamma_{L,\mu\nu}= \\
\nl = -\frac{1}{\sqrt{- \gamma}}\Lambda_0 ^\mu\left(H_{L\mu N}
Z'^N + \Lambda_1^\nu H_{L \mu\nu}\right) .\ea The quantities
$P_L$, $\mathcal{P}_L$ can be found as before, and now they are
\ba\label{mgcms} &&\left(\gamma^{01}g_{L N} - \frac{b_{LN}}{\sqrt{-
\gamma}}\right)Z'^N + \gamma^{0n}\Lambda^{\nu}_{n}g_{L\nu}-
\frac{\Lambda_1^\nu b_{L\nu}}{\sqrt{- \gamma}} =
-\frac{P_L}{T\sqrt{-\gamma}},\h \p_0 P_L=0,\\ \label{mgcq} &&
\gamma^{11}g_{LN}Z'^N + \gamma^{1n}\Lambda^{\nu}_{n}g_{L\nu} +
\frac{\Lambda_0 ^\nu b_{L\nu}}{\sqrt{- \gamma}} =
-\frac{\mathcal{P}_L}{T\sqrt{-\gamma}},\h \p_0 \mathcal{P}_L=0, \h
\p_1 \mathcal{P}_\mu=0.\ea The compatibility condition following
from the constraint (\ref{mga1s}) and from (\ref{mgcq}) coincides
with the previous one (\ref{ccs}).

As in the previous subsection, the equations (\ref{mgaems}) for
$L=\lambda$ lead to $\p_1\mathcal{P}_{\lambda}=0$. Consequently,
our next task is to consider the equations (\ref{mgaems}) for $L=a$
and the constraint (\ref{mga0s}). First of all, we will eliminate
the variables $Z'^\mu$ from them. To this end, we express $Z'^\mu$
through $Z'^a$ by using (\ref{mgcq}): \ba\label{mucs} Z'^\mu =
-\frac{\gamma^{1m}}{\gamma^{11}}\Lambda^\mu_m -
\left(g^{-1}\right)^{\mu\nu}\left[g_{\nu a}Z'^a +
\frac{1}{T\sqrt{-\gamma}\gamma^{11}}\left( \mathcal{P}_\nu +
T\Lambda_0^\rho b_{\nu \rho} \right)\right].\ea With the help of
(\ref{mucs}) and (\ref{ccs}), the equations (\ref{mgaems}) for $L=a$
and the constraint (\ref{mga0s}) acquire the form \ba\label{mgemfs}
&&h_{ab}Z''^b + \Gamma^{\bf{h}}_{a,bc}Z'^b Z'^c = \frac{1}{2}\p_a
\mathcal{U}^{\bf{g}} + 2\p_{[a}\mathcal{A}^{\bf{g}}_{b]}Z'^b,\\
\label{mgecfs} &&h_{ab}Z'^aZ'^b = \mathcal{U}^{\bf{g}}.\ea The
effective scalar and gauge potentials, expressed through the
background fields, are as follows \ba\nl
&&\mathcal{U}^{\bf{g}}=\frac{1}{\gamma\left(\gamma^{11}\right)^2}
\left[ \Lambda^{\mu}_{0}\Lambda^{\nu}_{0}g_{\mu\nu} +
\frac{1}{T^2} \left( \mathcal{P}_\mu + T\Lambda_0^\rho b_{\mu
\rho}\right)(g^{-1})^{\mu\nu} \left( \mathcal{P}_\nu +
T\Lambda_0^\lambda b_{\nu\lambda}\right)\right],
\\ \nl &&\mathcal{A}^{\bf{g}}_{a}= -
\frac{1}{T\sqrt{-\gamma}\gamma^{11}} \left[g_{a\mu}
(g^{-1})^{\mu\nu} \left(\mathcal{P}_\nu + T\Lambda _0 ^ \rho
b_{\nu\rho} \right)- T \Lambda_0 ^\rho b_{a \rho} \right] .\ea The
corresponding Lagrangian is  \ba\nl \mathcal{L}_{red}^{GA}(\sigma)
= -\frac{T}{2}\sqrt{- \gamma}\gamma^{11}\left(h_{ab}Z'^aZ'^b+
2\mathcal{A}^{\bf{g}}_a Z'^a + \mathcal{U}^{\bf{g}}\right) +
\frac{d}{d\s}\mathcal{P}_{\mu}\left(Z^\mu +
\Lambda_1^{\mu}\s\right).\ea

Since the equations (\ref{fems}), (\ref{ecs}) and (\ref{mgemfs}),
(\ref{mgecfs}) have the same form, for obtaining exact string
solutions, we  can proceed as before and use the derived formulas
after the replacements
$(g,\Gamma,\mathcal{U}^{s},\mathcal{A}^{s})$ $\to$
$(h,\Gamma^{\bf{h}}, \mathcal{U}^{\bf{g}},\mathcal{A}^{\bf{g}})$.
In particular, the solution depending on one of the coordinates
$X^a$ will be \ba\label{mgocss}\sigma\left(X^a\right)=\sigma_0 +
\int_{X_0^a}^{X^a}d x
\left(\frac{\mathcal{U}^{\bf{g}}}{h_{aa}}\right)^{-1/2}.\ea In
this case by integrating (\ref{mucs}), and replacing the solution
for $Z^\mu$ in the ansatz (\ref{sGA}), one obtains solution for
the string coordinates $X^\mu$ of the type $X^\mu(\tau, X^a)$: \ba
\label{Xs} &&X^\mu(\tau, X^a) = X_0^\mu + \Lambda_0 ^\mu \left[
\tau  - \frac{\gamma^{01}}{\gamma^{11}}
\sigma\left(X^a\right)\right] -\\ \nl && - \int_ {X_0^a}^ {X^a}
(g^{-1})^{\mu\nu} \left[g_{\nu a}+ \frac{\left(\mathcal{P}_\nu +
T\Lambda_0^\rho b_{\nu\rho}\right)} {T \sqrt{- \gamma}\gamma^{11}}
\left(\frac{\mathcal {U}^{\bf{g}}}{h_{aa}} \right)^{-1/2} \right]
d x .\ea To write down a solution of the type $X^\mu(\tau,\s)$,
one have to invert the solution (\ref{mgocss}):
$\sigma\left(X^a\right)\to X^a(\s)$. Then, $X^\mu(\tau,\s)$ are
given by \ba\label{aX} &&X^\mu(\tau, \s) = X_0^\mu + \Lambda_0
^\mu \left(\tau -\frac{\gamma^{01}}{\gamma^{11}} \sigma\right) -\\
\nl && - \int_{\s_0}^{\s} (g^{-1})^{\mu\nu}
\left[\frac{\left(\mathcal{P}_\nu + T\Lambda_0^\rho
b_{\nu\rho}\right)} {T \sqrt{- \gamma}\gamma^{11}} +g_{\nu a}
\left(\frac{\mathcal {U}^{\bf{g}}}{h_{aa}} \right)^{1/2}\right]
d\s .\ea

Let us also give the expression for $P_\mu$ after the elimination
of $Z'^\mu$ from (\ref{mgcms}) \ba\label{cmcq}
&&P_\mu(\s)=T\left[b_{\mu
a}-b_{\mu\nu}\left(g^{-1}\right)^{\nu\lambda}g_{\lambda a
}\right]Z'^a\\ \nl
&&+\frac{1}{\gamma^{11}}\left\{\gamma^{01}\mathcal{P}_{\mu} +
\frac{1}{\sqrt{-\gamma}}\left[T\Lambda^\nu_0g_{\mu\nu} -
b_{\mu\nu}\left(g^{-1}\right)^{\nu\lambda}\left(
\mathcal{P}_\lambda+T\Lambda^\rho_0b_{\lambda\rho}\right)\right]
\right\}.\ea These equalities connect the conserved momenta
$P_\mu$ with the constants of the motion $\mathcal{P}_{\mu}$.

To be able to take the tensionless limit $T\to 0$ in the above
formulas, we must use the $\lambda$-parametrization (\ref{tl}) of
$\gamma^{mn}$. The quantities, which depend on this
parametrization, and appear in the reduced equations of motion
and constraint (\ref{mgemfs}), (\ref{mgecfs}), and therefore - in
the solutions, are $\mathcal{U}^{\bf{g}}$ and
$\mathcal{A}^{\bf{g}}_{a}$. Now, they read \ba\nl
&&\mathcal{U}^{\bf{g}} =
-\frac{(2\lambda^0)^2}{\left[(2\lambda^0T)^2-(\lambda^1)^2\right]^2}
\left[ T^2\Lambda^{\mu}_{0}\Lambda^{\nu}_{0}g_{\mu\nu} + \left(
\mathcal{P}_\mu + T\Lambda_0^\rho b_{\mu
\rho}\right)(g^{-1})^{\mu\nu} \left( \mathcal{P}_\nu +
T\Lambda_0^\lambda b_{\nu\lambda}\right)\right],
\\ \nl &&\mathcal{A}^{\bf{g}}_{a}= -
\frac{2\lambda^0}{(2\lambda^0T)^2-(\lambda^1)^2} \left[g_{a\mu}
(g^{-1})^{\mu\nu} \left(\mathcal{P}_\nu + T\Lambda _0 ^ \rho
b_{\nu\rho} \right)- T \Lambda_0 ^\rho b_{a \rho} \right] .\ea If
one sets $\lambda^1=0$ and $2\lambda^0T=1$, the {\it conformal
gauge} results are obtained. If one puts $T=0$ in the above
equalities, they will correspond to {\it tensionless} strings. For
instance, the solution $X^\mu(\tau, X^a)$ reduces to \ba \nl
&&X^\mu(\tau, X^a)_{T=0} = X_0^\mu + \Lambda_0 ^\mu \left[ \tau +
\frac{\sigma\left(X^a\right)_{T=0}}{\lambda^1}\right] -\\ \nl && -
\int_ {X_0^a}^ {X^a} (g^{-1})^{\mu\nu} \left[g_{\nu a}-
\frac{2\lambda^0}{(\lambda^1)^2}\mathcal{P}_\nu
\left(\frac{\mathcal {U}^{\bf{g}}}{h_{aa}} \right)^{-1/2}_{T=0}
\right] d x .\ea

\vspace*{.5cm} {\bf Acknowledgments} \vspace*{.2cm}

This work is supported by the NSFB under contract $\Phi-1412/04$.




\end{document}